\documentclass[12pt]{article}
\usepackage{hyperref,amsfonts,amssymb, amsmath,mathrsfs}
\usepackage[english]{babel}

\def\nn{\nonumber }
\def\bq{ \begin{equation} }
\def\eq{ \end{equation} }
\def\ben{ \begin{eqnarray} }
\def\en{ \end{eqnarray} }

\def\on#1#2{\mathop{\vbox{\ialign{##\crcr\noalign{\kern2pt}
$\scriptstyle{#2}$\crcr\noalign{\kern2pt\nointerlineskip}
\kern-2pt$\hfil\displaystyle{#1}\hfil$\crcr}}}\limits}
\newtheorem{prop}{Theorem}

\begin{document}


\title{On a family of integrable systems on $S^2$ with a cubic integral of motion.}
\author{
A.V. Tsiganov\\
\\
\it\small St.Petersburg State University, St.Petersburg,
Russia}

\date{}
\maketitle

\begin{abstract}
We discuss a family of integrable systems on the sphere $S^2$
with an additional integral of third order in momenta. This
family contains the Coryachev-Chaplygin top, the Goryachev
system, the system recently discovered by Dullin and Matveev
and two new integrable systems. On the non-physical sphere
with zero radius all these systems are isomorphic to each
other.
\end{abstract}

\section{Introduction}
\setcounter{equation}{0}

Let us consider particle moving on the sphere $S^{2}=\{x\in
\mathbb R^3, |x|=a\}$. As coordinates on the phase space
$T^*S^2$ we choose entries of the vector $x=(x_1,x_2,x_3)$
and entries of the angular momentum vector $J=p\times x$. The
corresponding Poisson brackets reads as
\begin{equation}\label{e3}
\,\qquad \bigl\{J_i\,,J_j\,\bigr\}=\varepsilon_{ijk}J_k\,,
\qquad \bigl\{J_i\,,x_j\,\bigr\}=\varepsilon_{ijk}x_k \,,
\qquad \bigl\{x_i\,,x_j\,\bigr\}=0\,,
\end{equation}
where $\varepsilon_{ijk}$ is the totally skew-symmetric
tensor. The Casimir functions of the brackets (\ref{e3})
\bq\label{caz}
A=\sum_{i=1}^3 x_i^2=a^2,\qquad B=\sum_{i=1}^3 x_iJ_i=0\,,
\eq
are in the involution with any function on $T^*S^2$. So, for
the Liouville integrability of the corresponding equations of
motion it is enough to find only one additional integral of
motion, which is functionally independent of the Hamiltonian
$H$ and the Casimir functions.

If the corresponding Hamilton function $H$ has a natural
form, then according to Maupertuis's principle, integrable
system on $T^*S^2$ immediately gives a family of integrable
geodesic  on $S^2$. If the additional integral of this
integrable system is polynomial in momenta, integral of the
geodesic are also polynomial of the same degree.

In this note we discuss a family of integrable systems on
$T^*S^2$ with a cubic additional integral of motion. Among
such systems we distinguish the Goryachev-Chaplygin top
\cite{gor00,ch48} with the following integrals of motion
\bq\label{gor}
H=J_1^2+J_2^2+4J_3^2+cx_1\,,\qquad
K=2(J_1^2+J_2^2)J_3-cx_3J_1\,, \qquad c\in \mathbb R\,.
\eq
In \cite{ch48} Chaplygin found the separated variables
\bq\label{gor-var}
q_j=J_3\pm\sqrt{J_1^2+J_2^2+J_3^2},\qquad j=1,2,
\eq
in which dynamical equations are equal to
\bq\label{gor-eq}
(-1)^j\,(q_1-q_2)\dot{q_j}=2\sqrt{P(q_j)^2-a^2c^2q_j^2}\,,
\qquad P(\lambda)=\lambda^3-\lambda H+K\,.
\eq
These equations are reduced to the Abel-Jacobi equations and,
therefore, they are solved in quadratures \cite{ch48}.

In variables $q_j$ (\ref{gor-var}) integrals of motion
(\ref{gor-var}) read as
\bq\label{gor-int}
H=q_1^2+q_1q_2+q_2^2+cx_1,\qquad K=q_1q_2(q_1+q_2)-cx_3J_1\,.
\eq
In \cite{gor15} Goryachev de facto substituted special
generalizations of the variables $q_j$ (\ref{gor-var}) into
expressions similar to (\ref{gor-int}) in order to construct
new integrable system with a cubic integral of motion. In the
next section we generalize this result.

\section{A family of integrable systems on the sphere}
\setcounter{equation}{0}

Substituting canonical variables
\bq\label{genq}
{q}_j=\alpha J_3\pm\sqrt{J_1^2+J_2^2+f(x_3)J_3+g(x_3)},
\qquad \{q_1,q_2\}=0,
\eq
into the following ansatz for integrals of motion
\ben
H&=&q_1^2+q_1q_2+q_2^2+m(x_3)x_1,\nn\\
\label{int}\\
 K&=&q_1q_2(q_1+q_2)-n(x_3)J_1-\ell(x_3)x_1J_3\,,\nn
\en
one gets
\bq\label{H}
H=J_1^2+J_2^2+(3\alpha^2+f(x_3))J_3^2+m(x_3)x_1+g(x_3)\,,
\eq
and
\bq\label{K}
K= -2 \alpha J_3 \Bigl(-\alpha^2 J_3^2+J_1^2+J_2^2+f(x_3)
J_3^2 +g(x_3)\Bigr)-n(x_3) J_1-\ell(x_3) x_1 J_3\,.
\eq
Here $\alpha$ is an arbitrary numerical parameter, $f,g,m,n$
and $\ell$ are some functions of $x_3$ and of the single
non-trivial Casimir $a=\sqrt{x_1^2+x_2^2+x_3^2}$ (\ref{caz}).

\begin{prop}
On the phase space $T^*S^2$ functions $H$ (\ref{H}) and $K$
(\ref{K}) are in the involution with respect to the brackets
(\ref{e3}) if and only if function $n(x_3)$ is solution of
the following differential equation depending on $\alpha^2$
\ben
24\alpha^2-9&=&15\dfrac{x_3n' -n'' (a^2-x_3^2)}{n }
+\dfrac{3x_3n''-n'''(a^2-x_3^2)}{n'}\left(9-\dfrac{nn''}{n'^2}\right)\nn\\
\label{meq}\\
&+&n \left(
\dfrac{5x_3n'''-n''''(a^2-x_3^2)+3n''}{n'^2}\right).\nn
\en
 All another functions in
(\ref{H}-\ref{K}) are parameterized by $n(x_3)$
\ben
g(x_3)=\dfrac{d}{n(x_3)^2},\qquad
m(x_3)=-\dfrac{n'(x_3)}{\alpha},
\qquad \ell(x_3)= \dfrac{n(x_3)n''(x_3)}{n'(x_3)}, \nn\\
\label{fgml}\\
f(x_3)=
1-3\alpha^2-\alpha\,\dfrac{3x_3m(x_3)-2(a^2-x_3^2)m'(x_3)}{n(x_3)}
+\dfrac{x_3\ell(x_3) -(a^2-x_3^2)\ell'(x_3)}{n(x_3)}\,.\nn
\en
Here $d$ is arbitrary numerical parameter and $z'=\partial
z/\partial x_3$.
\end{prop}
The proof is straightforward.

In this note we consider particular solutions of differential
equation (\ref{meq}) only. Namely, substituting the following
ansatz
\bq\label{anz}
n(x_3)=c(x_3+e)^\beta\,,\qquad c,e,\beta\in\mathbb R\,,
\eq
in (\ref{meq}) one gets system of the algebraic equations on
the three parameters $\alpha,\beta$ and $e$ whereas two other
parameters $c$ and $d$ remain free.

\begin{prop}
Differential equation (\ref{meq}) has five particular
solutions in the form of (\ref{anz}) only:
\bq\label{msol}
\begin{array}{llll}
 1. &\pm\alpha=\beta=1,\quad & e=0,\qquad& n(x_3)=cx_3,\\
 \\
 2. &\pm\alpha=\beta=\dfrac{1}{3},\quad&e=0,\qquad &n(x_3)=cx_3^{1/3}, \\
 \\
 3. &\pm\alpha=\beta=\dfrac16,\quad &e=a, \qquad &n(x_3)=c(x_3+a)^{1/6},\\
 \\
 4. &\pm\alpha=\beta=\dfrac12,\quad & e\in \mathbb
 R,\qquad
 &n(x_3)=c(x_3+e)^{1/2}\,,\\
 \\
 5. &\pm\alpha=\beta =\dfrac14,\quad &e=a,\qquad
 &n(x_3)=c(x_3+a)^{1/4},.
\end{array}
\eq
The corresponding Hamilton functions (\ref{H}) are equal to
\ben
H_1&=&J_1^2+J_2^2+4J_3^2+cx_1+\dfrac{d}{x_3^2}\,,\nn\\
\nn\\
H_2&=&J_1^2+J_2^2+\dfrac{4}{3}J_3^2+\dfrac{cx_1}{x_3^{2/3}}+\dfrac{d}{x_3^{2/3}}\label{H5}\\
\nn\\
H_3&=&J_1^2+J_2^2+\left(\dfrac{7}{12}+\dfrac{x_3}{2(x_3+a)}\right)
J_3^2
+\dfrac{cx_1}{(x_3+a)^{5/6}}+\dfrac{d}{(x_3+a)^{1/3}}\,,\nn\\
\nn\\
H_4&=&J_1^2+J_2^2+\left(1+\dfrac{x_3}{x_3+e}-\dfrac{x_3^2-a^2}{4(x_3+e)^2}\right)
J_3^2+\dfrac{cx_1}{(x_3+e)^{1/2}}+\dfrac{d}{x_3+e}\,,\nn\\
\nn\\
H_5&=&J_1^2+J_2^2+\left(\dfrac{13}{16}+\dfrac{3x_3}{8(x_3+a)}\right)
J_3^2
+\dfrac{cx_1}{(x_3+a)^{3/4}}+\dfrac{d}{(x_3+a)^{1/2}}\,.\nn
\en
Transformation $\alpha\to -\alpha$ leads to the
transformation of the free parameters $(c,d)\to (-c,-d)$.
\end{prop}
Explicit expressions for additional cubic integrals of motion
$K_1,\ldots,K_5$ may be obtained by using definition
(\ref{K}) and equations (\ref{fgml},\ref{msol}).

The Hamilton function $H_1$ describes the Goryachev-Chaplygin
top \cite{ch48}. The second integrable system with
Hamiltonian $H_2$ was found by Goryachev \cite{gor15}. The
Hamilton function $H_4$ and the corresponding cubic integral
of motion $K_4$ was studied by Dullin and Matveev
\cite{dull04}. The third and fifth integrable systems with
Hamiltonians $H_3$ and $H_5$ are new.

At present we do not know whether our systems in implicit or
explicit forms (\ref{meq}-\ref{H5}) overlap with the families
of integrable geodesic flows on $S^2$ considered by
Selivanova \cite{sel99} and Kiyohara \cite{kiy01}. Recall
that in \cite{sel99,kiy01} all the geodesic flows are defined
in implicit form only (see also discussion in \cite{dull04}).

\section{The Lax matrices}
\setcounter{equation}{0}

In the fourth case (\ref{msol}-\ref{H5}) parameter $e$ is
free parameter and below we always put $e=0$.

Let us introduce $2\times 2$ hermitian matrix
\[
T(\lambda)=\left(%
\begin{array}{cc}
 A & B \\
 B^* & D \\
\end{array}%
\right)(\lambda),
\]
where $\lambda$ is a spectral parameter and
\ben
A(\lambda)&=&(\lambda-q_1)(\lambda-q_2)=\lambda^2-2\lambda\alpha
J_3
+\Bigl(\alpha^2-f(x_3)\Bigr)J_3^2-J_1^2-J_2^2-g(x_3)\,,\nn\\
\nn\\
B(\lambda)&=&(x_1+ix_2)m(x_3)\lambda+J_3(x_1+ix_2)\ell(x_3)+(J_1+i
J_2) n(x_3)\,,
\label{ABC}\\
\nn\\
 D(\lambda)&=& -n(x_3)^2\nn\,.
\en

The trace of this matrix
\[t(\lambda)=A(\lambda)+D(\lambda)=\lambda^2-\lambda H_L+K_L\]
gives rise integrals of motion in the involution for the
generalized Lagrange system
\[
H_L=2\alpha J_3\,,\qquad K_L=
\Bigl(\alpha^2-f(x_3)\Bigr)J_3^2-J_1^2-J_2^2-g(x_3)-n(x_3)^2\,.
\]
The corresponding equations of motion may be rewritten in the
form of the Lax triad
\[
\dfrac{d}{dt}T(\lambda)=\Bigl[T(\lambda),M(\lambda)\Bigr]+N(\lambda),\qquad
\mbox{\rm tr}\,N(\lambda)=0\,.
\]
In contrast with the Lax pair equations at $N(\lambda)=0$, in
the generic case determinant $\Delta(\lambda)=\det
T(\lambda)$ of the matrix $T(\lambda)$ (\ref{ABC}) is
dynamical function which do not commute with integrals of
motion:
\bq\label{G-det}
\begin{array}{ll}
 \beta=1\qquad& \Delta(\lambda)=-\dfrac{a^2}{\beta^2}\,\lambda^2\,
 \left(\dfrac{\partial n(x_3)}{\partial x_3}\right)^{2}+d,\\
 \\
 \beta=\dfrac13\qquad & \Delta(\lambda)=-\dfrac{a^2}{\beta^2}\,(\lambda+q_1+q_2)^2
 \,
 \left(\dfrac{\partial n(x_3)}{\partial x_3}\right)^{2}+d, \\
 \\
 \beta=\dfrac16\qquad & \Delta(\lambda)=-\dfrac{a}{\beta}\,(\lambda+q_1+q_2)^2\,
 \left(\dfrac{\partial n^2(x_3)}{\partial x_3}\right)+d,\\
 \\
 \beta=\dfrac12\qquad & \Delta(\lambda)=-\dfrac{a^2}{\beta^2}\,\lambda(\lambda+q_1+q_2)\,
 \left(\dfrac{\partial n(x_3)}{\partial x_3}\right)^{2}+d, \\
 \\
 \beta=\dfrac14\qquad &
 \Delta(\lambda)=-\dfrac{a}{\beta}\,\lambda(\lambda+q_1+q_2)\,
 \left(\dfrac{\partial n^2(x_3)}{\partial x_3}\right)+d\,.
\end{array}
\eq
At $\pm\alpha=\beta=1$ and $n(x_3)=cx_3$ matrix $T(\lambda)$
(\ref{ABC}) was constructed in \cite{kuzts1}. In this case
matrix $T(\lambda)$ defines representation of the Sklyanin
algebra on the space $T^*S^2$ associated with the symmetric
Neumann system \cite{kuzts1}.

\begin{prop}
If $n(x_3)$ is one of the particular solutions (\ref{msol})
of the differential equations (\ref{meq}) then $T(\lambda)$
(\ref{ABC}) satisfies to the following deformation of the
Sklyanin algebra
\bq
\{\,\on{T}{1}(\lambda),\,\on{T}{2}(\mu)\}= [r(\lambda-\mu),\,
\on{T}{1}(\lambda)\on{T}{2}(\mu)\,]+ Z(\lambda,\mu)\,,
\label{def-rrpoi}
\eq
where $\on{T}{1}(\lambda)= T(\lambda)\otimes
I\,,~\on{T}{2}(\mu)=I\otimes T(\mu)$, $I$ is a unit matrix
and
\bq
r(\lambda-\mu)=\dfrac{2i\alpha}{\lambda-\mu}\,\left(\begin{array}{cccc}
 1 & 0 & 0 & 0 \\
 0 & 0 & 1 & 0 \\
 0 & 1 & 0 & 0 \\
 0 & 0 & 0 & 1
\end{array}\right)\,.\label{rr}
\eq
Deformation $Z(\lambda,\mu)$ is hermitian matrix
\bq\label{S-mat}
Z(\lambda,\mu)=\left(\begin{array}{cccc}
 0 & u(\mu) & -u(\lambda) & 0 \\
 u^*(\mu) & 0 & w(\lambda,\mu) & 0 \\
 -u^*(\lambda) & w^*(\lambda,\mu) & 0 & 0 \\
 0 & 0 & 0 & 0
\end{array}\right),
\eq
which depends of the entries of $T(\lambda)$ (\ref{ABC})
only:
\begin{itemize}
 \item at $\beta=1$ we have $u=w=0$;
 \item at $\beta =\dfrac13,\dfrac16$ we have
\ben
&&u(\mu)=-4i\alpha\,\dfrac{\sqrt{\Delta(\mu)-d}}{D(\mu)}\,
\dfrac{\sqrt{\Delta(\mu)-d}\,B(\lambda)-\sqrt{\Delta(\lambda)-d}\,B(\mu)}{\lambda-\mu},
\nn\\
\nn\\
&&w(\lambda,\mu)=-4i\alpha\,\dfrac{\Delta(\lambda)-\Delta(\mu)}{\lambda-\mu}\,;\nn
\en
 \item at $\beta = \dfrac12,\dfrac14$ we have
\ben
&&u(\mu)=-2i\alpha\,\dfrac{\Delta(\mu)-d}{\mu\,D(\mu)}\,
\dfrac{\mu\,B(\lambda)-\lambda\,B(\mu)}{\lambda-\mu},\nn\\
\nn\\
&&w(\lambda,\mu)=-2i\alpha\,\dfrac{\Delta(\lambda)-\Delta(\mu)}{\lambda-\mu}\,.\nn
\en
\end{itemize}
Here
$\Delta(\lambda)=A(\lambda)\Delta(\lambda)-B(\lambda)B^*(\lambda)$
is determinant of the matrix $T(\lambda)$ (\ref{ABC}).

\end{prop}
The proof is straightforward.

One of the main properties of the Sklyanin algebra is that
for {\it any} numerical matrices ${\mathcal K}$ and for some
{\it special} dynamical matrices ${\mathcal K}$ coefficients
of the trace of matrix $\mathscr L(\lambda)={\mathcal
K}T(\lambda)$ give rise the commutative subalgebra
\[\{\mbox{\rm tr}\,{\mathcal K}T(\lambda)\,,\mbox{\rm tr}\,{\mathcal K}T(\mu)\}=0\,,\]
(see \cite{sokts2} and references within). All the generators
of this subalgebra are linear polynomials on coefficients of
entries $T_{ij}(\lambda)$, which are interpreted as integrals
of motion for integrable system associated with matrices
$T(\lambda)$ and $\mathcal K$ \cite{sokts2}.

Deformation of the Sklyanin algebra (\ref{def-rrpoi},
\ref{S-mat}) has the similar property.
\begin{prop}
If dynamical matrix $\mathcal K$ has the form
\[
\mathcal K=\left(%
\begin{array}{cc}
 \lambda+2\alpha J_3 & b_1 \\
 c_1 & 0 \\
\end{array}%
\right)\,\qquad b_1,c_1\in \mathbb C\,,
\]
then coefficients of the polynomial
\bq\label{P3}
P(\lambda)\equiv\mbox{\rm tr}\, {\mathcal K}T(\lambda)=
\lambda^3-\lambda\,H+K
\eq
are in the involution on $T^*S^2$.
\end{prop}
If $b_1=c_1=1/2$ then the first coefficient $H$ in
$P_3(\lambda)$ (\ref{P3}) coincides with one of the
Hamiltonians $H_1,\ldots,H_5$ (\ref{H5}) listed in the
Theorem 2, whereas the second coefficients $K$ is the
corresponding cubic integral $K_1,\ldots,K_5$ (\ref{K}). If
$b_1$ and $c_1$ is arbitrary one gets the same Hamiltonians
up to the suitable rescaling of $x$ and rotations
\bq
\label{rotE3} x\to b\, { U}\, x\,,\qquad J\to U J\,,
\eq
where $b$ is numerical parameter and $U$ is orthogonal
constant matrix.

The equations of motion associated with the Hamilton function
$H$ (\ref{P3}) may be rewritten as a Lax triad for the matrix
$\mathscr L(\lambda)=\mathcal KT(\lambda)$
\[
\dfrac{d}{dt}\mathscr L(\lambda)=\Bigl[\mathscr
L(\lambda),\mathscr M(\lambda)\Bigr]+\mathscr
N(\lambda),\qquad \mbox{\rm tr}\,\mathscr N(\lambda)=0\,.
\]
Here matrices $\mathscr M$ and $\mathscr N$ are restored from
the deformed algebra (\ref{def-rrpoi}) and definition of
Hamiltonian (\ref{P3}) in just the same way as for the usual
Sklyanin algebra \cite{kuzts1}.

\section{Isomorphism of the systems at $a=0$}

For all the considered systems (\ref{H5}) at $a=0$ additional
term $Z(\lambda,\mu)$ in (\ref{def-rrpoi}) is equal to zero
according to (\ref{G-det},\ref{S-mat})
\[ a=0\quad \Longrightarrow\quad\Delta(\lambda)=d \quad\Longrightarrow\quad
Z(\lambda,\mu)=0.
\]
In this case matrices $T(\lambda)$ associated with five
integrable systems (\ref{H5}) define five representations of
the Sklyanin algebra on the space $T^*S^2$. Of course, these
representations are related to each other by canonical
transformations.

\begin{prop}
At $a=0$, i.e. on the non-physical sphere $S^2$ with zero
radius, integrable systems listed in the Theorem 2 are
isomorphic to each other.
\end{prop}
To prove this Theorem we  introduce variables
\ben
p_j&=&\dfrac{1}{2\alpha i}\,\ln B(q_j)=\label{genp}\\
\nn\\
&=& \dfrac{1}{2\alpha i}\ln \Bigl(
\,q_j(x_1+ix_2)m(x_3)+J_3(x_1+ix_2)\ell(x_3)+(J_1+iJ_2)n(x_3)
\Bigr).\nn
\en
At $a=0$ variables $p_{1,2}$ and $q_{1,2}$ are canonical
Darboux variables according to (\ref{def-rrpoi})
\[
\{p_i,q_j\}=\delta_{ij}\,,\qquad
\{p_i,p_j\}=\{q_i,q_j\}=0\,,\qquad i,j=1,2.
\]
In order to construct canonical transformations which relate
integrable systems with Hamiltonians $H_1,\ldots,H_5$
(\ref{H5}) we have to identify variables $p_{1,2},q_{1,2}$
(\ref{genq},\ref{genp}) associated with the different
functions $n(x_3)$ (\ref{msol}).

We could not lift these symplectic transformations to the
Poisson maps. So, we can not assert that integrable systems
(\ref{H5}) are isomorphic on the generic symplectic leaves
(\ref{caz}).

The result of the Theorem 5 may be interpreted in the
following way. At $a=0$ on the special symplectic leaf of the
Lie algebra $e(3)$ there exists a germ of single integrable
system with Hamiltonian $H$ (\ref{gor}). Using canonical
symplectic transformations one can gets infinitely many
different forms of this integrable system. However, according
to the Theorem 2, these different forms of the germ admit
only a denumerable set of the continuation on the generic
symplectic leaves with conservation of the integrability
property. The similar observation for another family of
integrable systems on the sphere is discussed in \cite{ts03}.

\section{Summary}
Using the separation of variables for the Goryachev-Chaplygin
top we constructed a family of integrable systems on the
sphere with a cubic additional integral of motion. On the
non-physical sphere $S^2$ with zero radius these systems are
isomorphic to each other.

On this non-physical sphere  $a=0$ the separated variables
for all five systems coincide with the separated variables
for the Goryachev-Chaplygin top up to symplectic
transformations. It allows us to integrate equations of
motion in quadratures. On the usual sphere at $a\neq0$ the
separated variables are unknown. We suppose that these
variables may be constructed using the proposed deformation
of the Sklyanin algebra.

The research was partially supported by RFBR grant
02-01-00888.

\end{document}